\documentclass[aps,prl,twocolumn,superscriptaddress,preprintnumbers]{revtex4-1}%
\usepackage{epsf,epsfig}
\usepackage[utf8]{inputenc}
\usepackage{amssymb,amsmath,amsfonts}
\usepackage{color}
\usepackage{slashed}
\usepackage{tensor} 
\usepackage{braket}
\usepackage[colorlinks=true]{hyperref}  
\hypersetup{
    bookmarks=true,         
    unicode=false,          
    pdftoolbar=true,        
    pdfmenubar=true,        
    pdffitwindow=false,     
    pdfstartview={FitH},    
    colorlinks=true,       
    linkcolor=magenta, 
    citecolor=blue,        
    filecolor=magenta,      
    urlcolor=cyan           
} 

\newcommand{\bea}{\begin{eqnarray}}

\newcommand{\diff}{\mathrm{d}}
\newcommand{\eea}{\end{eqnarray}}
\usepackage{lipsum} 
\newcommand{\ba}{\begin{eqnarray}}
\newcommand{\ea}{\end{eqnarray}}

\newcommand{\beq}{\begin{equation}}
\newcommand{\eeq}{\end{equation}}
\newcommand{\beqa}{\begin{eqnarray}}
\newcommand{\eeqa}{\end{eqnarray}}
\newcommand{\beqar}{\begin{eqnarray*}}
\newcommand{\eeqar}{\end{eqnarray*}}

\newcommand{\ssc}{\scriptscriptstyle}
\newcommand{\eg}{{\it e.g.,}\ }
\newcommand{\ie}{{\it i.e.,}\ }

\usepackage{color}

\newcommand{\req}[1]{(\ref{#1})} 

\begin{document}

\title{A universal feature of charged entanglement entropy}
\author{Pablo Bueno}
\email{pablo.bueno-gomez@cern.ch}
\affiliation{CERN, Theoretical Physics Department,
CH-1211 Geneva 23, Switzerland}

\author{Pablo A. Cano}
\email{pabloantonio.cano@kuleuven.be}
\affiliation{Instituut voor Theoretische Fysica, KU Leuven. Celestijnenlaan 200D, B-3001 Leuven, Belgium}

\author{\'Angel Murcia}
\email{angel.murcia@csic.es}
\affiliation{Instituto de F\'isica Te\'orica UAM/CSIC,
	C/ Nicol\'as Cabrera, 13-15, C.U. Cantoblanco, 28049 Madrid, Spain}

\author{Alberto Rivadulla S\'anchez} 
\email{alberto.rivadulla.sanchez@usc.es}
\affiliation{Departamento de F\'isica de Part\'iculas, Universidade de Santiago de Compostela, E-15782 Santiago
de Compostela, Spain}
\affiliation{Instituto Galego de F\'isica de Altas Enerx\'ias (IGFAE), Universidade de Santiago de Compostela,
E-15782 Santiago de Compostela, Spain}


\preprint{IFT-UAM/CSIC-22-18}
\preprint{CERN-TH-2022-033}

\begin{abstract}
R\'enyi entropies, $S_n$, admit a natural generalization in the presence of global symmetries. These ``charged R\'enyi entropies'' are functions of the chemical potential $\mu$ conjugate to the charge contained in the entangling region  and reduce to the usual notions as $\mu\rightarrow 0$. For $n=1$, this provides a notion of charged entanglement entropy.   In this letter we prove that for a general $d (\geq 3)$-dimensional CFT, the leading correction to the uncharged entanglement entropy  across a spherical entangling surface is quadratic in the  chemical potential, positive definite, and universally controlled (up to fixed $d$-dependent constants) by the coefficients $C_J$ and $a_2$. These fully characterize, for a given theory, the current correlators $\braket{JJ}$ and $\braket{TJJ}$, as well as the energy flux measured at infinity produced by the insertion of the current operator. Our result is motivated by analytic holographic calculations for a special class of higher-curvature gravities coupled to a $(d-2)$-form in general dimensions as well as for free-fields in $d=4$. A proof for general theories and dimensions follows from previously known universal identities involving the magnetic response of twist operators introduced in \href{http://www.arXiv.org/abs/1310.4180}{{\tt arXiv:1310.4180}} and basic thermodynamic relations.
\end{abstract} 

\maketitle
The R\'enyi and entanglement entropies (EE) of spatial regions in the vacuum state of $d$-dimensional conformal field theories (CFTs) capture interesting universal information. This includes the Virasoro central charge $c$ for two-dimensional theories \cite{Calabrese:2004eu,Calabrese:2009qy}, the Euclidean partition function on the sphere in odd dimensions \cite{CHM,Dowker:2010yj}, the trace-anomaly coefficients in even dimensions \cite{Solodukhin:2008dh,Fursaev:2012mp,Safdi:2012sn,Miao:2015iba}, the stress-tensor two-point function charge $C_{\ssc T}$  \cite{Perlmutter:2013gua,Hung:2014npa,Faulkner:2015csl,Bueno1} and the thermal entropy coefficient $C_S$ \cite{Swingle:2013hga,Bueno3,HoloECG}, among others \cite{Lee:2014xwa,Lewkowycz:2014jia,Miao:2015dua}. From a different perspective, it has been in fact suggested that the full CFT data might be accessible from a long-distance expansion of the mutual/$N$-partite information \cite{Agon:2015ftl,Chen:2017hbk,Agon:2021lus,Agon:2021zvp,Casini:2021raa,Long:2016vkg,Chen:2016mya}. In this letter we consider a natural generalization of R\'enyi/entanglement entropies for theories with global symmetries \cite{Belin:2013uta} and add a new entry to the list of general relations satisfied by these quantities which connect them to various universal quantities.

Given a spatial bipartition, the (uncharged) R\'enyi entropy for some region $A$ is defined as $S_n\equiv \frac{1}{(1-n)} \log {\rm Tr} \rho_A^n$ where $\rho_A$ is the partial-trace density matrix associated to that region. The entanglement entropy $S_{\rm \ssc EE}$ is obtained as the $n\rightarrow 1$ limit of $S_n$. A charged notion of R\'enyi entropy was introduced in \cite{Belin:2013uta} for theories with global symmetries ---see also \cite{Lewkowycz:2013nqa,Wong:2013gua,Caputa:2013eka}.  This is given by
\begin{equation}
S_n(\mu)=\frac{1}{1-n} \log {\rm Tr} \left[\rho_A \frac{e^{\mu Q_A}}{n_A(\mu)} \right]^n \, ,
\end{equation}
where $Q_A$ is the total charge contained in the entangling region $A$, $\mu$ is the chemical potential conjugate to the charge and $n_A(\mu)$ a normalization factor. It is obvious from its definition that $S_n(\mu)$ reduces to $S_n$ as $\mu\rightarrow 0$. 
An interesting feature of  $S_n(\mu)$ is that, for spherical entangling surfaces, it admits a generalization of the conformal map of \cite{CHM,Hung:2011nu} which allows to evaluate this quantity from the ---usually simpler--- thermal entropy in the hyperbolic cylinder \cite{Belin:2013uta}. This enables one to perform explicit holographic and free-field calculations, which we exploit below. 
Additional studies of charged R\'enyi entropies and closely related notions can be found \eg in \cite{Belin:2013dva,Belin:2014mva,Pastras:2014oka,Nishioka:2013haa,Nishioka:2014mwa,Goldstein:2017bua,Tan:2019axb,Murciano:2019wdl,Bonsignori:2019naz}.

In the uncharged case, the EE universal term across a spherical entangling surface in a $d$-dimensional CFT reads ---see \eg \cite{Myers:2010tj,Myers:2010xs},
\begin{equation}
\label{eq:nuentro}
\frac{S_{\rm \ssc EE}}{ \nu_{d-1}}=a^{\star}\, , \quad \text{where} \quad \nu_{d-1}\equiv\begin{cases} (-)^{\frac{d-2}{2}} 4 \log(\frac{R}{\delta}) \, ,\\
(-)^{\frac{d-1}{2}} 2\pi \, ,\end{cases}
\end{equation}
respectively for even and odd $d$.  In this formula, $R$ is the radius of the spherical region and $\delta$ a UV regulator. 
In even dimensions, the universal quantity $a^\star$ is nothing but the A-type trace-anomaly coefficient \cite{Solodukhin:2008dh,Fursaev:2012mp,Safdi:2012sn,Miao:2015iba}. In odd $d$, $a^\star$ is proportional to the Euclidean partition function of the theory on the round sphere \cite{CHM,Dowker:2010yj}.

In this letter we show that the charged entanglement entropy for a spherical region is given, for general $d$-dimensional CFTs, by 
\begin{align}\label{Suniv}
\frac{S_{\rm \ssc EE}(\mu)}{\nu_{d-1}}&=a^{\star} + \frac{\pi^d C_J  }{(d-1)^2\Gamma(d-2)}\left[1+\frac{\displaystyle (d-2) a_2}{\displaystyle d(d-1)}\right] (\mu R)^2 
\end{align}
plus subleading corrections. Eq.\,\req{Suniv} can be alternatively formulated as a statement involving the first two derivatives of $S_n(\mu)$ with respect to $\mu$ evaluated at $n=1,\mu=0$ in an obvious way.
 In the above expression, $C_J$ and $a_2$ are two constants which characterize the corresponding CFT. On the one hand, $C_J$ is the only theory-dependent information which is not fully determined by conformal symmetry in the correlator of the current associated to the global symmetry, namely \cite{Osborn:1993cr}
\begin{equation}\label{CJJ}
\braket{J_a(x)J_b(0)}=\frac{ C_J}{|x|^{2(d-1)}}\left[\delta_{ab} - \frac{2x_a x_b}{|x|^2} \right]  \, . 
\end{equation}
As for $a_2$, its meaning can be understood from two different, albeit related, perspectives. On the one hand, 
the three-point function $\braket{TJJ}$  involves a complicated tensorial structure  ---which can be found in the appendix--- shared by all CFTs up to two theory-dependent coefficients \cite{Osborn:1993cr}. These coefficients can be chosen to be  $C_J$ and a second one denoted $a_2$. The latter can also be understood from conformal collider physics. Consider a CFT in flat space in its vacuum state and the insertion of a (smeared) current operator $\epsilon_a J^a$ for certain constant polarization tensor $\epsilon_a$. The expectation value of the energy flux measured at infinity in some direction $\vec n$ produced by such insertion is universally given by \cite{Hofman:2008ar}
\begin{equation}\label{Ea2}
\braket{\mathcal{E}(\vec{n})}_J=\frac{E}{\Omega_{(d-2)}} \left[1+a_2\left(\frac{|\epsilon\cdot n|^2}{|\epsilon|^2}-\frac{1}{d-1} \right) \right] \, ,
\end{equation}
where $\Omega_{(d-2)}$ is the volume of the unit radius $S^{(d-2)}$ and $E$ is the total energy. Again, the tensorial structure is fully fixed by symmetry, and all information about the corresponding CFT is in this case encoded in the coefficient $a_2$. Demanding the energy flux to be positive in all directions imposes the bounds \cite{Hofman:2008ar,Cano:2022ord} $-(d-1)/(d-2)\leq a_2\leq (d-1)$, which implies ---given the positivity of $C_J$ \cite{Osborn:1993cr}--- that the leading correction in Eq.\,\req{Suniv} is positive for general theories.

Formula \req{Suniv} then tells us that the charged entanglement entropy across a sphere of a general CFT for small values of the chemical potential has a leading correction to the uncharged result which is quadratic in the chemical potential, positive, and universally controlled by the charges $C_J,a_2$, which characterize the theory as explained above.



\textbf{Electromagnetic Quasitopological gravities.}
The realization that Eq.\,\req{Suniv} may be a universal relation came to us from holographic calculations, so we present those first. We consider the following  bulk theory for the metric field coupled to a $(d-2)$-form $B$ with field strength $H=\diff B$,
\begin{align} \label{EQf}
&I_{\rm EQG} =  \int  \frac{\diff^{d+1}x\sqrt{|g|} }{16 \pi G} \bigg[ R +\frac{d(d-1)}{L^2}-\frac{2H^2}{(d-1)!}\\ \notag
&+\frac{\lambda L^2 \mathcal{X}_{4}}{(d-2)(d-3)}+\frac{2 \alpha_1 L^2}{(d-1)!}\mathcal{L}_{RH^2}^{(1)}+\frac{2 \alpha_2 L^2}{(d-1)!}\mathcal{L}_{RH^2}^{(2)} \bigg]\, ,
\end{align}
where $G$ is the Newton constant, $L$ is a length scale, $\lambda,\alpha_1,\alpha_2$ are dimensionless couplings,
$\mathcal{X}_{4}$ is the Gauss-Bonnet density, and\footnote{Note that also at fourth order in derivatives we could have included a term of the type $H^4$, which also belongs to the EQTG class. However, since we are interested in the regime of small charge, that term plays no role in our discussion and we have simply omitted it.}
\begin{equation}
\begin{aligned}
\mathcal{L}_{RH^2}^{(1)}&\equiv H^2R-(d-1)(2d-1)\tensor{R}{^{\mu\nu}_{\rho\sigma}}\tensor{(H^2)}{^{\rho\sigma}_{\mu\nu}}\, ,\\
\mathcal{L}_{RH^2}^{(2)}&\equiv \tensor{R}{^{\mu}_{\nu}} \tensor{(H^2)}{^{\nu}_{\mu}}-(d-1)\tensor{R}{^{\mu\nu}_{\rho\sigma}}\tensor{(H^2)}{^{\rho\sigma}_{\mu\nu}}\, ,
\end{aligned}
\end{equation}
with $\left(  H^2 \right)^{\rho \sigma}{}_{\mu \nu}\equiv H^{\rho \sigma \alpha_3 \alpha_4 \dots \alpha_{d-1}} H_{\mu \nu \alpha_3 \alpha_4 \dots \alpha_{d-1}} $
are two Electromagnetic Quasitopological theories (EQGs) \cite{Cano:2020qhy,Cano:2022ord}. These belong to a class of modifications of Einstein gravity with distinct properties, including simple black hole solutions and linearized spectrum, analytic thermodynamics, as well as providing a basis for general-order effective actions \cite{Quasi2,Quasi,Dehghani:2011vu,Ahmed:2017jod,Cisterna:2017umf,PabloPablo,Hennigar:2016gkm,PabloPablo2,Hennigar:2017ego,Bueno:2019ltp,Bueno:2019ycr}. From an AdS/CFT perspective \cite{Maldacena,Witten,Gubser}, Eq.\,\req{EQf} defines models of $(d-1)$-dimensional CFTs parametrized by the bulk action couplings. Different CFT magnitudes will involve different functions of those  couplings \cite{Buchel:2009sk,Myers:2010jv,HoloECG,Mir:2019ecg,Camanho:2009hu}, which can be used to elucidate universal patterns when some of those magnitudes  in fact display the same dependence. This approach has been successfully used before, \eg in  \cite{Myers:2010tj,Myers:2010xs,Bueno1,Bueno2,Bueno:2018yzo,Mezei:2014zla,Chu:2016tps,Li:2018drw}.

Eq.\,\req{EQf} can be mapped to a different theory with a vector field by dualizing the $B$-field. The field strength of the dual vector field $F=\diff A$, is then identified as 
$
F=4\pi G \ell_{*}^{-1} (d-1)! \star \left[ \partial\mathcal{L}/\partial H\right]\, , 
$
where $\ell_{*}$ is an undetermined length scale that we introduce so that $A_{\mu}$ has units of energy.  The bulk gauge field 
 $A_{\mu}$ is holographically dual to the current $J^{a}$ of a  global U$(1)$. The parameters $C_{J}$ and $a_2$ associated to $J^a$ were determined in \cite{Cano:2022ord}, finding
 \begin{align}\label{eq:CJgen}
C_{J}^{\rm EQG}&=\frac{\Gamma(d)}{4 \pi^{d/2 + 1} \Gamma(d/2 - 1) } \frac{\ell_{*}^2\tilde L^{d-3}}{\alpha_{\rm eff} G}\, ,\\ 
a_2^{\rm EQG} &=  - \frac{2 d (d-1) \left[ (2d-1) \alpha_1 + \alpha_2 \right] f_\infty}{(d-2)\alpha_{\rm eff}}\, ,
\label{eq:a2}
\end{align}
where 
\begin{equation}\label{alphae}
\alpha_{\rm eff}\equiv 1-f_{\infty}\alpha_{1}(3d^2-7d+2)-f_{\infty}\alpha_{2}(d-2)\, ,
\end{equation}
 $f_{\infty}\equiv  L^2/\tilde L^2$ and $\tilde L$ is the AdS$_{(d+1)}$ radius.

Now, the (charged) R\'enyi entropy across a spherical entangling surface of radius $R$ in the vacuum state can be obtained, on general grounds, from the thermal entropy on $\mathbb{S}^1_{2\pi R}\times \mathbb{H}^{d-1}_R$ \cite{CHM,Hung:2011nu,Belin:2013uta}. In the holographic context, the calculation amounts to computing the thermal entropy of an AdS$_{(d+1)}$ hyperbolic black hole charged under the gauge field at a temperature $T_0=1/(2\pi R)$. For our theory \req{EQf}, this takes the form
\begin{equation}
\begin{aligned}
\diff s^2&=\frac{-L^2}{f_{\infty}R^2}\left[\frac{r^2}{L^2}f-1\right]\diff t^2+\frac{\diff r^2}{\frac{r^2}{L^2}f-1}+r^2\diff \Xi^2\, ,\\ 
H&=Q \, \omega_{\mathbb{H}^{d-1}}\, ,
\end{aligned}
\end{equation}
where $\diff \Xi^2$ is the metric of the unit hyperbolic space $\mathbb{H}_1^{d-1}$ and $\omega_{\mathbb{H}^{d-1}}$ its volume form. The factor $\frac{L^2}{f_{\infty}R^2}$ has been introduced so that the boundary metric is conformal to $\diff s^2_{\mathbb{S}^1\times \mathbb{H}^{d-1}}=-\diff t^2+R^2\diff \Xi^2$. The equations of motion for $f(r)$ and its explicit form can be found in the appendix. The temperature of the black holes can be written as
\begin{align}\label{temp}
 &T=\frac{T_0}{2 x \sqrt{f_\infty}    \left(1 - 2 p^2 \alpha_1 - 2   \lambda x^{-2} \right) } \Bigg[  x^2 d-(d-2)     \\ \notag
 &+  \frac{(d-4) \lambda}{ x^{2}}  - \frac{2 p^2}{(d-1)} \left[x^2 - d   (3(d-1)\alpha_1 + \alpha_2) \right]  \Bigg] \, ,
 \end{align}
where we introduced $x\equiv r_{+}/L$, $p\equiv Q L r_{+}^{-d+1}$ and $r_+$ is the outer horizon position. We also need the value of the chemical potential of the boundary theory. This is nothing but the asymptotic value of the electrostatic potential $A_{t}$ at $r\rightarrow\infty$, which is fixed by the condition that $A_{t}|_{r_{+}}=0$. We find
\begin{equation}
\begin{aligned}\label{mue}
&\mu=  \frac{L p}{\ell_{*} \sqrt{f_{\infty}} R } \Bigg[ \frac{x}{(d-2)} \\ &  - \frac{\alpha_1}{x} \left( 3 (d-1) + \frac{T}{T_0}2 x  \sqrt{f_{\infty}}  \right) - \frac{ \alpha_2 }{x}  \Bigg] \, .
\end{aligned}
\end{equation}
Finally, we need the Wald entropy \cite{Wald:1993nt,Iyer:1994ys} of the solutions. We obtain
\begin{equation}
	S = \frac{x^{d-1}L^{d-1} V_{\mathbb{H}^{d-1}}}{4 G} \left[ 1 + 2p^2\alpha_1- \frac{2 (d-1) \lambda}{(d-3) x^2} \right]\, ,
	\label{eq:entropy1}
\end{equation}
where $V_{\mathbb{H}^{d-1}}\equiv \nu_{d-1}\Omega_{d-1}/(4\pi)$ is the regularized volume of the unit hyperbolic space.
As explained earlier, this computes the holographic charged entanglement entropy when $T=T_0$. Observe that in the above expression, the dependence on $\mu$ appears through $x$ and $p$, so we would need to obtain $x(\mu)$ and $p(\mu)$ from Eqs.\,\req{temp} and \req{mue} evaluated for such temperature in order to obtain an explicit formula for $S^{\rm  univ}_{\rm \ssc EE}(\mu)$. This cannot be done explicitly for arbitrary values of $\mu$, but it is possible for small values of $\mu R$. The result for the first two orders reads
\begin{widetext}
\begin{equation}
\frac{S^{\rm EQG}_{\rm \ssc EE}(\mu)}{\nu_{d-1}}=a^{\star}_{\rm GB} +\frac{ \pi^{(d-2)/2} (d-2)^2 [1-3d(d-1)\alpha_1f_{\infty}-d \alpha_2 f_{\infty}]}{(d-1) 8\Gamma(d/2) \alpha_{\mathrm{eff}}^2} \frac{\tilde L^{d-3}\ell_{*}^2}{G} (\mu R)^2+\mathcal{O}(\mu^4)\, ,
\end{equation}
\end{widetext}
where $ \alpha_{\mathrm{eff}}$ was defined in \req{alphae}. Now, the constant term is the $a^{\star}$ charge for our EQG theory, which reduces to the Gauss-Bonnet gravity one, as terms involving the $B$ form do not contribute to it. Explicitly, this reads  \cite{Myers:2010tj}
\begin{equation}
a^{\star}_{\rm GB} =\frac{\tilde L^{d-1}}{8G} \frac{\pi^{(d-2)/2}}{\Gamma(d/2)}\left[1-\frac{2(d-1)}{d-3}\lambda f_{\infty}\right]\,.
\label{eq:astargb}
\end{equation}
As mentioned earlier, this is the expected result for the (uncharged) entanglement entropy across a spherical surface in $d$ dimensions.
Now, the leading correction coming from the chemical potential has a complicated non-polynomial dependence on the gravitational couplings $\alpha_1,\alpha_2$. However, this conspires to produce a linear combination of the charges $C_J^{\rm EQG}$ and $C_J^{\rm EQG} \cdot a_2^{\rm EQG}$. Indeed, using Eqs.\,\req{eq:CJgen} it is easy to see that the above formula reduces to Eq.\,\req{Suniv}. In the appendix we show that  Eq.\,\req{Suniv} in fact holds for an infinite family of EQGs of general orders. 

The fact that $S^{\rm  univ}_{\rm \ssc EE}(\mu)$ takes this simple form for such a large family of holographic theories leads us to think that this may actually be a relation which holds for completely general CFTs. Before proving that this is indeed the case, we can perform an additional check in a completely different context.



\textbf{Free fields.}
The result for the charged R\'enyi entropy associated to global phase rotations for a Dirac fermion and a scalar field in $d=4$  has been computed in \cite{Belin:2013uta} using heat-kernel techniques. We review these calculations in the appendix, where we also fix a typo in the Dirac fermion result reported in that paper.  The correct results read, respectively,
\begin{align}\label{snfs}
S^{\rm f}_n&=\frac{\nu_3}{24} \left[ \frac{(1+n)(7+37n^2)}{120 n^3}+\frac{(1+n) (\mu R)^2}{n}\right] \, , \\ \notag 
S^{\rm s}_n&=\frac{\nu_3}{24} \left[\frac{(1+n)(1+n^2)}{60 n^3}+\frac{(1+n)(\mu R)^2}{ 2n}+ |\mu R|^3 \right]\, .
\end{align}
Interestingly, the exact dependence on $\mu $ is much simpler than for our holographic theories, for which, as we saw earlier, a completely explicit formula cannot be obtained.  
It is then straightforward to obtain the result of interest for the entanglement entropy expansion. One finds
\begin{equation}
\begin{aligned}\label{seefs}
\frac{S^{\rm f}_{\rm \ssc EE}(\mu)}{\nu_3 } &=a^{\star}_{ \rm f}+\frac{(\mu R)^2}{12} \, , \\   \frac{S^{\rm s}_{\rm \ssc EE}(\mu)}{\nu_3 }&=a^{\star}_{ \rm s}+\frac{(\mu R)^2}{24}+\frac{|\mu R|^3}{24}\, ,
\end{aligned}
\end{equation}
where
$
a^{\star}_{\rm f}=11/360$, $a^{\star}_{\rm s}=1/360\, , 
$
are the trace-anomaly coefficients corresponding to a Dirac fermion and a real scalar field, respectively \cite{Casini:2010kt,Dowker:2010bu,Lee:2014zaa}. Now, the values of the charges $C_J$ and $a_2$ for these two models are also well-known and read \cite{Osborn:1993cr,Petkou:1994ad,Hofman:2008ar,Chowdhury:2012km}
\begin{equation}
\begin{aligned}
C_J^{\rm f}&=\frac{1}{\pi^4}\, , \quad  C_J^{\rm s}=\frac{1}{4\pi^4}\, ,\\
a_2^{\rm f }&=-\frac{3}{2}\, , \quad  a_2^{\rm s}=3\, .
\end{aligned}
\end{equation}
It is then straightforward to verify that Eqs.\,\req{seefs} satisfy the relation \req{Suniv}.


\textbf{General CFTs.}
The previous results strongly suggest that Eq.\,\req{Suniv} holds for general CFTs. As it turns out, a proof of such universality can be easily achieved using a combination of the results presented in Ref.~\cite{Belin:2013uta} along with some thermodynamic identities. In order to do this, we need to depart momentarily from the vacuum temperature $T_0$ and consider a CFT on the hyperbolic cylinder at an arbitrary temperature $T$. The thermal entropy of a given CFT in such state can be used to compute the R\'enyi entropy $S_n(\mu)$ across a spherical entangling region \cite{Hung:2011nu,Belin:2013uta}, the R\'enyi index being related to the temperature by $n=T_0/T$. 

In order to proceed, we need to consider a set of related quantities: 
the twist operators $\sigma_n(\mu)$. In the Replica trick approach to the evaluation of R\'enyi/entanglement entropy, the entangling region is cut from each of the spacetime copies and consecutive copies are sewn together along the entangling surface. Such boundary conditions can be understood as produced by the insertion of $(d-2)$-dimensional operators along the entangling surface \cite{Calabrese:2004eu,Calabrese:2009qy,Hung:2011nu,Hung:2014npa}. In the charged R\'enyi/EE case, the entangling surface carries  a ``magnetic flux''  $-i n \mu$ which can be understood as attaching a Dirac sheet to the twist operators \cite{Belin:2013uta}.


 The leading divergence in the correlator of $\sigma_n(\mu)$ with the current operator defines the so called ``magnetic response'' $k_{n}(\mu)$ as \cite{Belin:2013uta}
\begin{equation}
\left\langle J_a \sigma_n(\mu) \right\rangle = \frac{{\rm i} k_n(\mu)}{2 \pi} \frac{\epsilon_{ab}n^b}{y^{d-1}}\, ,
\end{equation}
where $y$ is the distance between the insertions, $n^b$ is a unit vector normal to $J_a$ from the twist operator insertion and $\epsilon_{ab}$ is the volume form of the two-dimensional space  orthogonal to the entangling surface. 
In the case of a spherical entangling surface, the magnetic response is given by \cite{Belin:2013uta}
\begin{equation}
k_n(\mu) = 2\pi n R^{d-1} \rho(n, \mu)\, ,
\label{eq:kndef}
\end{equation}
where $ \rho(n, \mu)$ is the charge density of the CFT on the hyperbolic cylinder at temperature $T=T_0/n$. As it turns out, this quantity has a universal expansion around $n=1$ and $\mu=0$ whose leading terms can be expressed in terms of the coefficients characterizing the $\braket{TJJ}$ correlator. We have \cite{Belin:2013uta} 
\begin{equation}
\begin{aligned}
\label{k01}
k_n\Big|_{n=1,\mu=0}&=
\partial_{n} k_{n}\Big|_{n=1,\mu=0}=0\, , \\
\partial_{\mu} k_{n}\Big|_{n=1,\mu=0}&=\frac{16 R \pi^{d+1}}{\Gamma(d+1)}[\hat c+\hat e]\, , \\
 \partial_n\partial_{\mu} k_{n}\Big|_{n=1,\mu=0}&=\frac{16 R \pi^{d+1}}{d \Gamma(d+1)} \left[ 2 \hat{c} - d (d-3) \hat{e} \right]\, ,
\end{aligned}
\end{equation}
where the charges $\hat c,\hat e$ are related to $C_J,a_2$ by \cite{Chowdhury:2012km,Belin:2013uta}
\begin{equation}
\begin{aligned}\label{ceCJ}
\hat c&=\frac{C_{J}(d-2)\Gamma\left(\frac{d+2}{2}\right)}{2\pi^{d/2}(d-1)^3}\left[d(d-1)-a_2\right]\, ,\\
\hat e&=\frac{C_{J}\Gamma\left(\frac{d+2}{2}\right)}{2\pi^{d/2}(d-1)^3}\left[d-1+(d-2)a_2\right]\, .
\end{aligned}
\end{equation}

Let us now consider the thermal entropy $S$ of the CFT on the hyperbolic cylinder. In the grand canonical ensemble, the first law of thermodynamics reads
\begin{equation}
\diff \Omega=-S \diff T-N \diff \mu\, ,
\end{equation}
where $\Omega$ is the grand potential and $N=V_{\mathbb{H}^{d-1}}R^{d-1}\rho$ is the total charge. 
From the first law the following thermodynamic relation can be obtained
\begin{equation}
\partial_{\mu}S=-\partial_{\mu}\partial_{T}\Omega=-\partial_{T}\partial_{\mu}\Omega=\partial_{T}N\, .
\end{equation}
Writing $N$ in terms of the magnetic response $k_{n}(\mu)$, and using that $\partial_{T}=-\frac{T_0}{T^2}\partial_{n}$, we have
\begin{equation}\label{pmuS}
\partial_{\mu}S=-\frac{T_0  V_{\mathbb{H}^{d-1}} }{2\pi T^2} \partial_n \left(\frac{k_n(\mu)}{n} \right)\, .
\end{equation}
Expanding the derivatives, evaluating for $n=1$ $(T=T_0)$ and $\mu=0$ and using Eqs.\,\req{k01}, it immediately follows that the first derivative term vanishes, \ie
\begin{equation}
\partial_{\mu}S_{\rm \ssc EE} \big|_{\mu=0}=0\, .
\end{equation}
Taking a second derivative with respect to $\mu$ in Eq.\,\req{pmuS}, we have
\begin{equation}
\partial^2_{\mu}S=-\frac{T_0 V_{\mathbb{H}^{d-1}}}{2\pi T^2}\partial_{\mu}\partial_{n}\left(\frac{k_{n}(\mu)}{n}\right)\, .
\end{equation}
Evaluating again for $n=1$ $(T=T_0)$ and $\mu=0$, we have
\begin{equation}
\partial^2_{\mu}S_{\rm \ssc EE}\big|_{\mu=0}=R V_{\mathbb{H}^{d-1}}\left[\partial_{\mu}k_{n}-\partial_{\mu}\partial_{n}k_{n}\right]\Big|_{n=1,\mu=0}\, .
\end{equation}
Using then Eq.~\req{k01}, we can rewrite this as
\begin{equation}
\partial^2_{\mu}S_{\rm \ssc EE}\big|_{\mu=0}=V_{\mathbb{H}^{d-1}} \frac{16 (d-2) R^2 \pi^{d+1}}{d\Gamma(d+1)}\left[\hat c+d \hat e\right]\, ,
\label{eq:univrel}
\end{equation}
which, via Eq.\,\req{ceCJ} reduces to Eq.\,\req{Suniv}. This therefore completes the proof that such relation is universally valid for arbitrary CFTs.

{\bf Final comments.} Our formula \req{Suniv} holds for general CFTs in $d\geq 3$. In $d=2$, there are various reasons to expect a different situation. On the one hand, observe that the coefficient $a_2$ is not even defined in that case. Similarly, from Eq.\,\req{eq:CJgen} it is clear that $C_J$ for our holographic calculations is divergent for $d=2$ and therefore meaningless. The free-field results reported in \cite{Belin:2013uta} also suggest a different structure in that case, including possible linear terms in $\mu$ or jumps in $S_n(\mu)$ as $n$ and $\mu$ vary. It would be interesting to investigate these features further ---natural candidates would be three-dimensional holographic EQGs \cite{Bueno:2021krl}.

On a different front, it would also be interesting to rederive Eq.\,\req{Suniv} using the techniques developed in \cite{Rosenhaus:2014woa}. In the case of a small perturbation by a relevant operator $\mathcal{O}$, the leading correction to the EE across a sphere was shown to be quadratic in the perturbation and proportional to a double integral of  $\braket{K \mathcal{O}\mathcal{O}}-\braket{\mathcal{O}\mathcal{O}}$, where $K$ is the modular Hamiltonian of $\rho_A$ ---which for spheres involves an integral of the stress tensor. In the present context, it would be natural to relate  $\mathcal{O}$ to the charge operator, which would bring about integrals of $\braket{TJJ}$ and $\braket{JJ}$, precisely as expected from Eq.\,\req{Suniv}.

In \cite{Magan:2021myk}, a somewhat similar universal relation for charged R\'enyi entropies ---involving the uncharged result plus an extra term--- was obtained in the case of discrete symmetry groups. It would be nice to study the connection between Eq.\,\req{Suniv} and the approach developed in that paper and \cite{Casini:2019kex} in the case of continuous groups.

A particularly interesting application of our formula is to the case of supersymmetric (S) CFTs, which come with a global $R$-symmetry group. For instance, for $d=4$, $\mathcal{N}=1$ SCFTs one has a U$(1)_{R}$ current with \cite{Osborn:1998qu,Barnes:2005bm,Hofman:2008ar}
\begin{equation}
C_{J}^{\mathcal{N}=1,\, {\rm U}(1)_R}=\frac{4 c}{\pi^4}\, ,\quad a_2^{\mathcal{N}=1,\,{\rm U}(1)_R}=3\left(1-\frac{a}{c}\right)\, ,
\end{equation}
and therefore, our formula \req{Suniv} yields the prediction 
\begin{equation}
S_{\rm \ssc EE}^{\mathcal{N}=1,\,{\rm U}(1)_R}=\nu_{3}\left[a+\frac{2}{3}\left(c-\frac{a}{3} \right)(\mu R)^2+\dots\right]\, ,
\end{equation}
where we used $a^{\star}=a$ and $c$ is the other trace-anomaly coefficient. Similarly, for  $\mathcal{N}=2$ SCFTs, the $R$-symmetry group is U$(1)_{R}\times$ SU$(2)_{R}$. Using the corresponding values of $C_J$ and $a_2$ 
\cite{Shapere:2008zf,Hofman:2009ug}, one finds\footnote{In the  ${\rm SU}(2)_R$ case one should understand that $\mu$ couples to a U$(1)$ subgroup of it. }
\begin{equation}
\begin{aligned}
S_{\rm \ssc EE}^{\mathcal{N}=2,\,{\rm U}(1)_R}&=\nu_{3}\left[a+2\left(c-\frac{a}{3} \right)(\mu R)^2+\dots\right]\, , \\
S_{\rm \ssc EE}^{\mathcal{N}=2,\,{\rm SU}(2)_R}&=\nu_{3}\left[a+\frac{1}{6}\left(2c-a \right)(\mu R)^2+\dots\right]\, .
\end{aligned}
\end{equation}
 It would be interesting to verify these predictions using alternative methods.

Finally,  it is natural to wonder what additional relations connecting quantum information measures and universal CFT quantities  may  still remain to be discovered.


\vskip1mm
\noindent
\textbf{Acknowledgements}
\vskip1mm

\noindent
We would like to thank Nikolay Bobev and Javier Mag\'an for useful discussions.
The work of P.A.C. is supported by a postdoctoral
fellowship from the Research Foundation - Flanders (FWO grant 12ZH121N). The work of \'A.M. is funded by the Spanish FPU Grant No. FPU17/04964. \'A.M. is further supported by the MCIU/AEI/FEDER UE grant PGC2018-095205-B-I00 and by the Spanish Research
Agency (Agencia Estatal de Investigaci\'on) through the Grant IFT Centro de Excelencia
Severo Ochoa No CEX2020-001007-S, funded by MCIN/AEI/10.13039/501100011033. A.R.S.
is supported by the Spanish MECD grant FPU18/03719. The work of A.R.S. is further funded
by AEI-Spain (under project PID2020-114157GB-I00 and Unidad de Excelencia Mar\'ia de
Maeztu MDM-2016-0692), by Xunta de Galicia-Conseller\'ia de Educaci\'on (Centro singular de
investigaci\'on de Galicia accreditation 2019-2022, and project ED431C-2021/14), and by the
European Union FEDER.
%

\onecolumngrid  \vspace{1cm} 
\begin{center}  
{\Large\bf Appendices} 
\end{center} 
\appendix 

\section{Formulas and conventions for the $\braket{TJJ}$ three-point function  }\label{TJJ1}
The flat-space three-point function $\braket{TJJ}$ for a general CFT is given by \cite{Osborn:1993cr,Erdmenger:1996yc}
\begin{equation}
	\braket{T_{ab}(x_1)J_c(x_2)J_d(x_3)}=\frac{t_{abef}(X_{23}) I^e_{c}(x_{21}) I^f_{d}(x_{31})}{|x_{12}|^d|x_{13}|^d |x_{23}|^{d-2}} \, .
\end{equation}
Here, $I_{ab}(x)$ is the same tensorial structure that appears in the definition of the current correlator, \ie
\begin{equation}
	I_{ab} \equiv \delta_{ab} - 2 \frac{x_a x_b}{|x|^2} \, ,
\end{equation}
while $t_{abcd}(X_{23})$ is given by
\begin{equation}
	   \begin{aligned}
		t_{abcd}(X^a) &\equiv \hat a h^{(1)}_{ab}(\hat{X}^a)\delta_{cd} + \hat b h^{(1)}_{ab}(\hat{X}^a)h^{(1)}_{cd}(\hat{X}^a) + \hat c h^{(2)}_{abcd}(\hat{X}^a) + \hat e h^{(3)}_{abcd}(\hat{X}^a) \, , \\
		h^{(1)}_{ab}(\hat{X}^a) &\equiv \hat{X}_a\hat{X}_b - \frac{1}{d}\delta_{ab} \, , \\
		h^{(2)}_{abcd}(\hat{X}^a) &\equiv 4\tensor{\hat{X}}{_{(a}}\tensor{\delta}{_{b)(d}}\tensor{\hat{X}}{_{c)}}-\frac{4}{d}\hat{X}_a\hat{X}_b \delta_{cd} - \frac{4}{d}\hat{X}_c\hat{X}_d\delta_{ab} + \frac{4}{d^2}\delta_{ab}\delta_{cd} \, , \\
		h^{(3)}_{abcd} &\equiv \delta_{ac}\delta_{bd} + \delta_{ad}\delta_{bc} - \frac{2}{d} \delta_{ab}\delta_{cd} \, . \\
	\end{aligned}
	\label{quantitiesTJJ}
\end{equation}
Notice that indices are raised and lowered with the flat metric $g_{ab} = \delta_{ab}$. Also, we have defined
\begin{equation}
	x_{12}^a \equiv x_1^a - x_2^a \, ,\quad X_{12}^a \equiv \frac{x_{13}^a}{|x_{13}|^2} - \frac{x_{23}^a}{|x_{23}|^2} \, ,\quad \hat{X}_{23}^a \equiv \frac{X_{23}^a}{|X_{23}|}\, ,
\end{equation}
and similarly for the different permutations of the indices. The expression for $t_{abcd}(X^a)$ in \eqref{quantitiesTJJ} depends on four parameters, $\hat{a}$, $\hat{b}$, $\hat{c}$ and $\hat{e}$. However, imposing conservation of the currents it is found that they are related as \cite{Osborn:1993cr}
\begin{equation}
	d \hat{a} - 2 \hat{b} + 2 (d-2) \hat{c} = 0 \, , \quad \hat{b} - d (d-2) \hat{e} = 0 \, ,
\end{equation}
so only two of them are independent, which we choose to be $\hat{c}$ and $\hat{e}$. Studying Ward identities that relate three- and two-point functions, it was found in Ref. \cite{Osborn:1993cr} that these two constants are related to the central charge $C_J$ defined in Eq.\,\req{CJJ} as
\begin{equation}
	C_J = \frac{2 \pi^{d/2}}{\Gamma(d / 2 + 1)} \left( \hat{c} + \hat{e} \right) \, .
\end{equation}
Finally, these parameters can also be related to the coefficient $a_2$ appearing in the energy flux Eq.\,\req{Ea2}, since this quantity is proportional to a component of the correlator $\braket{TJJ}$. This has been done explicitly in \cite{Chowdhury:2012km}, finding
\begin{equation}
	a_2 = \frac{(d-1) ( d (d-2) \hat{e} - \hat{c} )}{(d-2) ( \hat{c} + \hat{e} )} \, .
\end{equation}
Therefore, these relations allow us to express the three-point function $\braket{TJJ}$, which depends on two parameters, entirely in terms of $C_J$ and $a_2$.

\section{Free field calculations}
\label{densis}
In this appendix we summarize the calculation of the charged R\'enyi entropies for free scalars and fermions in $d=4$ using heat-kernel techniques \cite{Vassilevich:2003xt,https://doi.org/10.1112/S0024609398004780,Grigoryan1994,Camporesi:1994ga}. Our results here closely follow the derivation in \cite{Belin:2013uta}, but we use the opportunity to correct a few typos that appear in that paper, which include the final expression for $S_n(\mu)$ in the case of the free fermion.


We will compute the charged R\'enyi entropy from the free energy on $\mathbb{S}^1\times \mathbb{H}^3$. In order to do that, we will use the heat-kernel on such space. For product spaces this factorizes, so one has
\begin{equation}
K_{\mathbb{S}^1\times \mathbb{H}^3}=K_{\mathbb{S}^1}(\theta_1,\theta_2,t) K_{\mathbb{H}^3}(\vec y_1,\vec y_2,t)\, .
\end{equation}
Following \cite{Belin:2013uta}, we consider a purely imaginary chemical potential for a global U$(1)$ charge associated to phase rotations of the fields. This is related to the real chemical potential we use throughout the rest of the paper by $\mu_{\rm \ssc E}= 2\pi {\rm i} R \mu$. Incorporating the chemical potential in the heat-kernel amounts to requiring this to satisfy an appropriate boundary condition. This reads
\begin{equation}
K_{\mathbb{S}^1}(\theta_1+2\pi n,\theta_2,t)= (-)^f {\rm e}^{-{\rm i}n \mu_{\rm\ssc E}}K_{\mathbb{S}^1}(\theta_1,\theta_2,t)\, ,
\end{equation}
where $f=1$ for Dirac fermions and $f=0$ for scalars. This  is achieved by a modified disk heat-kernel of the form\footnote{In \cite{Belin:2013uta} there is a missing ``$n$'' that should be multiplying ``$2\pi m$'' in Eq.\,(A.6) and which propagates throughout the whole appendix.}
\begin{equation}
K_{\mathbb{S}^1}(\theta_1,\theta_2,t)=\frac{1}{\sqrt{4\pi t}} \sum_{m\in \mathbb{Z}} {\rm e}^{-\frac{(\theta_2-\theta_1+2\pi n m)^2}{4t}} {\rm e}^{-{\rm i}m (n \mu_{\rm \ssc E}+\pi f)}\, .
\end{equation}
Indeed, upon substitution of $\theta_1\rightarrow \theta_1+2\pi n$, the numerator of the exponent of the first term becomes $(\theta_2-\theta_1+2\pi n (m-1))^2$. Since the sum is over all integers, one can shift the index $m= m'+1$ leaving the first term as it was originally and producing an overall $ (-)^f {\rm e}^{-{\rm i}n \mu_{\rm\ssc E}}$ from the second term.
The equal-point heat kernel then reads
\begin{equation}
K_{\mathbb{S}^1}(0,0,t)= \frac{1}{\sqrt{4\pi t}} \sum_{m\in \mathbb{Z}} {\rm e}^{-\frac{\pi^2 n^2 m^2}{t}}{\rm e}^{-{\rm i}m (n \mu_{\rm \ssc E}+\pi f) }\, .
\end{equation}
On the other hand, the  equal-point heat kernel for the hyperbolic space reads \cite{Belin:2013uta}
\begin{equation}
K_{\mathbb{H}^3}(0,0,t)=\frac{(1+3f)}{(4\pi t)^{3/2}} \left[1+\frac{t f}{2} \right]\, .
\end{equation}
From these, the free energy on $\mathbb{S}^1_{(2\pi n)}\times \mathbb{H}^3$ can be computed as
\begin{align}
F_n(\mu_{\rm\ssc E})&=\frac{(-)^{f+1}}{2}V_{\mathbb{H}^3} (2\pi n) \int_0^{\infty} \frac{\diff t}{t}  K_{\mathbb{S}^1}(0,0,t)K_{\mathbb{H}^3}(0,0,t)\, , \\
&=(-)^{f+1}\frac{n(1+3f) }{16\pi} V_{\mathbb{H}^3}\sum_{m\in \mathbb{Z}}   {\rm e}^{- {\rm i} m (n \mu_{\rm \ssc E} +\pi f)} \int_0^{\infty}\frac{\diff t}{t^3} \left[1+\frac{tf}{2} \right]{\rm e}^{-\frac{\pi^2 n^2 m^2}{t}}\, .
\end{align}
The zero mode in the disk heat kernel gives rise to a divergence in the free energy \cite{Belin:2013uta}, so we can ignore it and get for the regulated free energy
\begin{align}
F_n
&=(-)^{f+1}\frac{(1+3f) V_{\mathbb{H}^3}}{16\pi^5 n^3}  \sum_{m\in \mathbb{Z}^+}   (-)^{m f}  \cos[m  n \mu_{\rm \ssc E}] \frac{(2+f m^2 n^2 \pi^2)}{m^4}\, , \\ &= (-)^{f+1}\frac{(1+3f) V_{\mathbb{H}^3}}{16\pi^5 n^3}  \left[ \frac{f n^2\pi^2}{2} \left( {\rm Li}_2[(-)^f {\rm e}^{-{\rm i} n \mu_{\rm \ssc E}}] +  {\rm Li}_2[(-)^f {\rm e}^{{\rm i} n \mu_{\rm \ssc E}}]  \right)+ {\rm Li}_4[(-)^f {\rm e}^{-{\rm i } n \mu_{\rm \ssc E}}] +  {\rm Li}_4[(-)^f {\rm e}^{{\rm i} n \mu_{\rm \ssc E}}]  \right]\, .
\end{align}
From this, the charged R\'enyi entropy can be obtained as
\begin{equation}
S_n(\mu_{\rm \ssc E})= \frac{1}{n-1} \left[ F_n(\mu_{\rm \ssc E}) - n  F_1(\mu_{\rm \ssc E}) \right]\, .
\end{equation}
We find, respectively, for the Dirac fermion and the free scalar,
\begin{align}
S^{\rm f}_n(\mu_{\rm \ssc E})&=\frac{V_{\mathbb{H}^3}}{48 \pi} \left[\frac{(1+n)(7+37n^2)}{30 n^3}-\frac{(1+n)\mu_{\rm \ssc E}^2}{ n \pi^2} \right] \, , \quad
S^{\rm s}_n(\mu_{\rm \ssc E})&=\frac{V_{\mathbb{H}^3}}{48 \pi} \left[\frac{(1+n)(1+n^2)}{15 n^3}-\frac{(1+n)\mu_{\rm \ssc E}^2}{ 2n \pi^2} +  \frac{|\mu_{\rm \ssc E}|^3}{2\pi^3} \right]\, .
\end{align}
The scalar formula agrees with the one presented in \cite{Belin:2013uta}, but the fermion one is different. There seems to be a missing $1/(4\pi^2)$ multiplying $\mu_{\rm \ssc E}^2$ in their Eq.\,(A.25).
Finally, writing these in terms of $\nu_3$ and $\mu$, we find the formulas presented in the main text in Eq.\,\req{snfs}, and from those the entanglement entropy expansions appearing in Eq.\,\req{seefs}, which agree with our general formula \req{Suniv}.

\section{Charged Entanglement Entropy in EQGs}\label{app:eqgsee}
 
In this appendix we verify the universal relation Eq.\,\eqref{Suniv} for an infinite set of higher-derivative theories  of gravity coupled to a $(d-2)$ form, which are of arbitrary order in the curvature and in the gauge field strength. These theories generalize those presented in Eq.\,\eqref{EQf} and were discovered and studied in \cite{Cano:2022ord}. Their action reads
\begin{equation}
\begin{aligned}
\label{eq:geneqg}
I_{\rm EQG}^{\rm gen} =\frac{1}{16 \pi G}\int \mathrm{d}^{d+1} x \sqrt{\vert g \vert} \bigg [&R+\frac{d(d-1)}{L^2}- \frac{2H^2}{(d-1)!}+\frac{\lambda L^2 \mathcal{X}_4 }{(d-2)(d-3)} \\  & +\frac{2}{(d-1)!}\sum_{s=0}^\infty \sum_{m=1}^\infty L^{2(s+m-1)}\left ( \alpha_{1,s,m} \mathcal{L}^{(a)}_{d,s,m}+\alpha_{2,s,m} \mathcal{L}^{(b)}_{d,s,m}  \right) \bigg ]\, ,
\end{aligned} 
\end{equation}
where
\begin{equation}
\begin{aligned}
\mathcal{X}_4\equiv &+R^2-4R_{\mu\nu}R^{\mu\nu}+R_{\mu\nu\rho\sigma}R^{\mu\nu\rho\sigma}\, , \\
\mathcal{L}^{(a)}_{d,s,m}\equiv &\left(s R\tensor{\left(R^{s-1}\right)}{^{\mu\nu}_{\rho\sigma}}+\kappa_{d,s,m} \tensor{\left(R^{s}\right)}{^{\mu\nu}_{\rho\sigma}}+2s (s-1) \tensor{R}{_{\gamma}^{\mu}} \tensor{R}{^{\beta}_{\rho}}\tensor{\left(R^{s-2}\right)}{^{\gamma\nu}_{\beta\sigma}} \right) (H^2)_{\mu\nu}{}^{\rho\sigma}(H^2)^{m-1}\, ,
\label{eq:egqa}\\
\mathcal{L}^{(b)}_{d,s,m}\equiv &\frac{1}{2}\left(2s\tensor{R}{_{\mu}^{\alpha}}\tensor{\delta}{_{\nu}^{\beta}}+g_{d,s,m}\tensor{R}{^{\alpha\beta}_{\mu\nu}}\right)\tensor{\left(R^{s-1}\right)}{^{\mu\nu}_{\rho\sigma}}(H^2)^{\rho\sigma}{}_{\alpha\beta} (H^2)^{m-1}\, ,\\
\end{aligned}
\end{equation}
and where we used the notation
\begin{equation}
\begin{aligned}
\left(  H^2 \right)^{\rho \sigma}{}_{\mu \nu}\equiv H^{\rho \sigma \alpha_3 \alpha_4 \dots \alpha_{d-1}} H_{\mu \nu \alpha_3 \alpha_4 \dots \alpha_{d-1}} \, , \quad \tensor{\left(R^{k}\right)}{^{\mu\nu}_{\rho\sigma}}\equiv R^{\mu \nu}{}_{\alpha_1 \beta_1} R^{\alpha_1 \beta_1}{}_{\alpha_2 \beta_2} \dots R^{\alpha_{k-1} \beta_{k-1}}{}_{\rho \sigma}\,,
\end{aligned}
\end{equation}
$g_{d,s,m}\equiv -d(s-1)-2(d-1)m$ and $\kappa_{d,s,m}\equiv (1-g_{d,s,m})g_{d,s,m}/2$. 

Any members of the infinite family of theories captured by Eq.\,\eqref{eq:geneqg} are examples of Electromagnetic Quasitopological Gravities \cite{Cano:2020qhy}. By taking the only non-vanishing couplings to be $\alpha_{1,1,1} \equiv\alpha_1$ and $\alpha_{2,1,1} \equiv \alpha_2$ we recover the theories discussed in the main text, defined in Eq.\,\eqref{EQf}. The theories \eqref{eq:geneqg} admit charged black-hole solutions with spherical, planar or hyperbolic sections. Their computational treatment is fairly similar and, since here we are interested only in solutions with hyperbolic sections, we restrict ourselves to this case. Such solutions are of the form
\begin{equation}\label{hyperbolicBHapp}
\diff s^2=\frac{-L^2}{f_\infty R^2}\left[\frac{r^2}{L^2}f-1\right]\diff t^2+\frac{\diff r^2}{\left[\frac{r^2}{L^2}f-1\right]}+r^2\diff \Xi^2\, ,\quad 
H=Q \, \omega_{\mathbb{H}^{d-1}}\, ,
\end{equation}
where $\diff \Xi^2$ is the metric of the unit hyperbolic space $\mathbb{H}_1^{d-1}$, $\omega_{\mathbb{H}^{d-1}} $ the associated volume form and $f_\infty\equiv L^2/\tilde L^2$ ---where $\tilde L$ is the AdS$_{(d+1)}$ radius---  can be written in terms of the Gauss-Bonnet coupling as $2 \lambda f_\infty=1-\sqrt{1-4 \lambda}$. Remarkably, the above solutions are characterized by a single metric function $f(r)$. The full non-linear equations of \eqref{eq:geneqg} collapse to a single first-order differential equation for $f(r)$ which can be integrated once, yielding the following algebraic equation
\begin{equation}\label{eqfgeneral}
\begin{aligned}
0=&+\frac{r^2}{L^2} (1-f)- \frac{16 \pi R  \sqrt{f_\infty} G M}{ (d-1) L V_{\mathbb{H}^{d-1}} r^{d-2}}+\frac{2 Q^2}{(d-2)(d-1) r^{2(d-2)}}+ \frac{\lambda r^2}{L^2} f^2\\ &+ \sum_{s,m} \frac{Q^{2m} L^{2m} (-2)^s \Gamma(d)^{m-1}}{r^{2m(d-1)}} f^{s-1}  \bigg[ - \frac{2s}{d-1}((1-2m)(d-1)+1-ds) \alpha_{1,s,m}+\frac{s \alpha_{2,s,m} }{d-1}\\ &- \bigg (\bigg (1-2m-4s+4ms +\frac{2s(ds-1)}{d-1}\bigg ) \alpha_{1,s,m}+ \frac{(s-1) \alpha_{2,s,m}}{d-1}\bigg ) \frac{r^2}{L^2} f \bigg]\,.
\end{aligned}
\end{equation}
Here $M$ is an integration constant to be identified with the mass of the solution and $\sum_{s,m}\equiv \sum_{s=0}^\infty \sum_{m=1}^\infty$. In the special case of the four-derivative theory \eqref{EQf}, the equation of motion for $f(r)$ is simplified to
\begin{align}
0=&\frac{r^2}{L^2} (1-f)- \frac{16 \pi R  \sqrt{f_\infty} G M}{ (d-1) L V_{\mathbb{H}^{d-1}} r^{d-2}}+\frac{2 Q^2}{(d-2)(d-1) r^{2(d-2)}}+ \frac{\lambda r^2}{L^2} f^2 -2\frac{Q^{2} L^{2}}{r^{2(d-1)}}   \bigg[4 \alpha_{1}+ \frac{\alpha_{2} }{d-1}-  \alpha_{1} \frac{r^2}{L^2} f \bigg]\,.
\end{align}
This is a quadratic equation for $f$ which can be easily solved. Choosing the solution that reduces to the Einstein gravity result in the limit $\lambda \rightarrow 0$, we find
\begin{equation}
\begin{aligned}
f(r)&=\frac{\mathcal{B}(r)- \sqrt{\mathcal{B}^2(r)-4 \lambda \mathcal{C}(r)}}{2 \lambda}\, , \\ \mathcal{B}(r)&\equiv 1-\frac{2 L^2 Q^2 \alpha_1}{r^{2(d-1)}}\, , \quad \mathcal{C}(r)\equiv 1+\frac{2 Q^2 L^4}{r^{2d}} \bigg[\frac{r^2}{L^2(d-1)(d-2)}-4\alpha_1-\frac{\alpha_2}{(d-1)} \bigg]-\frac{16 \pi G M L R \sqrt{f_\infty} }{(d-1) V_{\mathbb{H}_1^{d-1}} r^d } \, .
\end{aligned}
\end{equation}
Let us now go back to the generic case of theories with any number of derivatives. Assume that $g_{tt}$ has some zero along the positive real axis and let $r_+=\mathrm{max}\left\lbrace r \in \mathbb{R}^+\vert f(r)=\frac{L^2}{r^2}\right\rbrace$. Defining $x \equiv r_+/L$ and $ p \equiv Q L^{2-d} x^{1-d}$, and evaluating \req{eqfgeneral} at $r=r_+$, the mass $M$ of the subsequent black hole solution can be seen to to be
\begin{equation}
\begin{aligned}
\left[\frac{16 \pi R \sqrt{f_\infty} G}{L^{d-1} V_{\mathbb{H}^{d-1}}}\right] M = & +(d-1)x^{d-2} \left ( x^2-1\right ) + \frac{2 p^2 x^d}{(d-2)}+ (d-1) \lambda x^{d-4} \\ & + \sum_{s,m}\frac{(-2)^s \Gamma(d)^{m-1} p^{2m} }{x^{2s-d}} (-(d-1)(1-2m-2s)\alpha_{1,s,m}+\alpha_{2,s,m})\, .
\end{aligned}
\end{equation}
Similarly, taking into account that the temperature $T$ is given by $4 \pi R \sqrt{f_\infty} T=\dfrac{r_+^2}{L} f'(r_+)+\dfrac{2 L}{r_+}$, we find the following expression,
\begin{equation}
\begin{aligned}
&4 \pi R \sqrt{f_\infty} T=\frac{(d-1)(2+d(x^2-1)+(d-4) \lambda x^{-2})-2 p^2 x^{2}}{(d-1)(x- 2\lambda x^{-1})+\sum_{s,m} (-2)^s  \Gamma(d)^{m-1} p^{2m} s   (2s+d(2m-1)-2m-1) x^{3-2s} \alpha_{1,s,m}}\\  & -\frac{\sum_{s,m} (-2)^s \Gamma(d)^{m-1} p^{2m} x^{-2(s-1)} (2s-2m+d(2m-1))((2m+2s-1)\alpha_{1,s,m}+(d-1)^{-1}\alpha_{2,s,m})}{x- 2\lambda x^{-1}+\sum_{s,m}(-2)^s  \Gamma(d)^{m-1} (d-1)^{-1} p^{2m} s   (2s+d(2m-1)-2m-1) x^{3-2s} \alpha_{1,s,m}}  \,.
\end{aligned}
\end{equation}
The computation of the black hole entropy $S$ is carried out  using the Iyer-Wald formula \cite{Wald:1993nt,Iyer:1994ys},
\begin{equation}
S=-2\pi\int_{\Sigma} \diff^{d-1}x\sqrt{h}\frac{\partial\mathcal{L}}{\partial R_{\mu\nu\rho\sigma}}\epsilon_{\mu\nu}\epsilon_{\rho\sigma}\, ,
\end{equation}
where $\Sigma$ is a cross section of the black hole horizon and $\epsilon_{\mu\nu}$ is its binormal. 
This can be straightforwardly applied to the theories \req{eq:geneqg} and further evaluated for the black hole metric \req{hyperbolicBHapp}, yielding
\begin{equation}
\label{eq:entrogen}
S=\frac{x^{d-1} L^{d-1} V_{\mathbb{H}^{d-1}}}{4G} \bigg[ 1-\frac{2 (d-1) \lambda}{(d-3)x^2}- \sum_{s,m} \frac{(-2)^s \Gamma(d)^{m-1}s  p^{2m} }{x^{2(s-1)}} \alpha_{1,s,m}\bigg] \, .
\end{equation}
Interestingly, the entropy does not receive any corrections from the density $\mathcal{L}^{(b)}_{d,s,m}$ but only from $\mathcal{X}_4$ and $\mathcal{L}^{(a)}_{d,s,m}$.  

As explained in the main text, the chemical potential $\mu$ is defined as the asymptotic value of the electrostatic potential $A_t$  after demanding that $\left. A_t  \right \vert_{r_+}=0$. Such electrostatic potential is the only active component of the dual vector field $A$ in the case of magnetic configurations and is given by
\begin{equation}
F=4 \pi G \ell_\ast^{-1} (d-1)! \star \frac{\partial \mathcal{L}}{\partial H}\, , \quad F= \diff A\,, \quad A=A_t \diff t\,.
\end{equation}
It is illustrative to show the expression for $A_t$ in the particular case of the four-derivative theory \eqref{EQf}. We find
\begin{equation}
\ell_\ast A_t(r)=-\frac{Q L}{R \sqrt{f_\infty} r^{d-2}} \bigg[ \frac{1}{d-2}+\alpha_1((-3d+1) f-r f')-\alpha_2 f \bigg] + \ell_\ast A_{t,\infty}\,,
\end{equation}
where $A_{t,\infty}$ represents the asymptotic value for $A_t$ chosen so that $\left. A_t  \right \vert_{r_+}=0$, and hence $\mu=A_{t,\infty}$. 

In the general case of theories with an arbitrary number of derivatives, it can be checked that the chemical potential $\mu$ reads
\begin{equation}
\left[\frac{L^{d-2} x^{d-1} V_{\mathbb{H}^{d-1}}  \ell_\ast}{4 \pi G}  \right] \mu=\frac{\partial M}{\partial p}-T \frac{\partial S}{\partial p}\,.
\end{equation}
Taking into account this expression and the previous presented thermodynamic magnitudes, it is possible to show that the first law of black hole thermodynamics holds, namely,
\begin{equation}
\mathrm{d}M=T \mathrm{d}S + \mu\, \mathrm{d}N \, , \quad \text{where} \quad N\equiv Q \cdot  \left[ \frac{ V_{\mathbb{H}^{d-1}}  \ell_\ast}{4 \pi G}  \right]
\end{equation}
is the total charge in the boundary theory.

Now, our goal is to compute the vacuum charged EE for the boundary theory across a spherical entangling surface of radius $R$. Such entanglement entropy can be obtained from the thermal entropy of the same theory placed on the hyperbolic cylinder $\mathbb{S}^{1} \times \mathbb{H}^{d-1}_R$ at temperature $T_0=1/(2\pi R)$ \cite{CHM,Hung:2011nu,Belin:2013uta}. Then, using the holographic dictionary, such thermal entropy turns out to be just the Wald entropy of a black hole with hyperbolic horizon, \ie
\begin{equation}
S_{\rm \ssc EE}(\mu)=S(T_0,\mu)\,.
\end{equation}
Consequently, for the derivation of the charged entanglement entropy, we need to evaluate the Wald entropy \eqref{eq:entrogen} at temperature $T=T_0$ and in terms of the chemical potential $\mu$. Above, in Eq.\,\eqref{eq:entrogen} we wrote $S=S(x,p)$, so we need to find the inverse functions $x=x(T_0,\mu)$ and $p=p(T_0,\mu)$. We will carry out this procedure in a perturbative fashion in $\mu$ and we will restrict ourselves to the leading-order corrections (so that it suffices to keep only the terms quadratic in $H$). We find
\begin{equation}
\begin{aligned}
x&=\hat{x}+\delta x_2 (l_\ast \mu)^2+ \mathcal{O}(\mu^4)\, , \quad p= \delta p_1 (l_\ast \mu) +\mathcal{O}(\mu^3)\, , \quad \hat{x}=\frac{1}{\sqrt{f_\infty}} \, , \\ \delta p_1&= \frac{2 f_\infty R}{L\left ( \frac{2}{d-2}+\sum_{s=0}^\infty(-2 f_\infty)^s ((d+2d s-1)\alpha_{1,s}+\alpha_{2,s}\right) } \, , \\ \delta x_2&= -\frac{(\delta p_1)^2 (2+\sum_{s=0}^\infty (-2)^s f_\infty^s (2-4s+d(d-3+2(d-1)s+4s^2) \alpha_{1,s}+(d-2+2s)\alpha_{2,s}) )}{2(d-1)^2 (f_\infty -2) \sqrt{f_\infty} }\,.
\end{aligned}
\end{equation}
Plugging the (perturbative) expressions found above into Eq.\,\eqref{eq:entrogen}, we find that the entanglement entropy to quadratic order in $\mu$ reads
\begin{equation}
S_{\rm \ssc EE}=\frac{\tilde{L}^{d-1}V_{\mathbb{H}^{d-1}} }{4  G} \bigg[1-\frac{2(d-1)}{d-3}\lambda f_{\infty}+ \left ( \frac{\sqrt{f_\infty} R}{L}\right)^2 \frac{(l_\ast \mu)^2}{\alpha_{\rm eff}} \left (\frac{(d-2)^2}{d-1}+ \frac{(d-2)^2 \beta_{\rm eff}}{(d-1)^2 \alpha_{\rm eff} } \right) \bigg]+\mathcal{O}(\mu^4)\,,
\end{equation}
where we have defined the parameters
\begin{equation}
\begin{aligned}
\alpha_{\rm eff}&\equiv 1+ \sum_{s=0}^\infty (-2)^{s-1} f_\infty^s (2-d) \left ( (d-1+2ds) \alpha_{1,s}+\alpha_{2,s} \right)\,, \\
\beta_{\rm eff}&\equiv \sum_{s=0}^\infty  (-2 f_\infty)^s (d-1) s \left ( (2ds-1) \alpha_{1,s}+\alpha_{2,s}\right)\,.\label{betaeff}
\end{aligned}
\end{equation}

Our next goal will be trying to express the charged entanglement entropy (up to quadratic order in $\mu^2$) in terms of the charges $C_J$ and $a_2$ of the CFT dual to the theories \eqref{eq:geneqg}. On the one hand, if $F$ denotes the dual field strength of $H$, $C_J$ is obtained by working out the effective gauge coupling of $F^2$ when we evaluate the action on a pure AdS background \cite{Belin:2013uta}. Owing the fact that we will restrict ourselves to quadratic terms in $\mu$, it is enough to keep in the action \eqref{eq:geneqg} those terms up to quadratic order in $H$. If $I_{\rm dual}^{F^2}$ denotes the dual theory to any theory containing terms of up to second-order in $H$, then according to \cite{Cano:2021tfs,Cano:2021hje,Cano:2022ord} $I_{\rm dual}^{F^2}$  can be shown to be
\begin{align}
\label{eq:dualth}
I_{\rm dual}^{F^2}=\int \frac{\mathrm{d}^{d+1} x \sqrt{\vert g \vert}}{16 \pi G}& \bigg [R+\frac{d(d-1)}{L^2}+\frac{\lambda L^2 \mathcal{X}_4 }{(d-2)(d-3)} -( \tilde{Q}^{-1}) _{\mu \nu}{}^{\rho \sigma}F^{\mu \nu} F_{\rho \sigma} \bigg] \, ,
\end{align}
where
\begin{equation}
\begin{aligned}
\tilde{Q}^{\mu \nu}{}_{\rho \sigma}&\equiv  \frac{12}{(d-1)(d-2)} Q^{[\alpha \beta}{}_{\alpha \beta} \delta^{\mu \nu]}{}_{\rho \sigma}\,, \\
Q^{\alpha \beta}{}_{\rho \sigma}&\equiv \delta^{\alpha\beta}{}_{\rho\sigma} -\sum_{s=0}^\infty  \bigg[ \frac{1}{2}  \tensor{\left(R^{s-1}\right)}{^{\mu\nu}_{\rho\sigma}} \left(2s\tensor{R}{_{\mu}^{[\alpha}}\tensor{\delta}{_{\nu}^{\beta]}}+g_{d,s,1}\tensor{R}{^{\alpha\beta}_{\mu\nu}}\right)\alpha_{2,s}\\&  \hspace{1cm} +\left(s R\tensor{\left(R^{s-1}\right)}{^{\alpha \beta}_{\rho\sigma}}+\kappa_{d,s,1} \tensor{\left(R^{s}\right)}{^{\alpha\beta}_{\rho\sigma}}+2s (s-1) \tensor{\left(R^{s-2}\right)}{^{\mu[\alpha}_{\nu[\rho|}} \tensor{R}{_{\mu}^{\beta]}} \tensor{R}{^{\nu}_{|\sigma]}}  \right) \alpha_{1,s}\bigg]\, ,
\end{aligned}
\end{equation}
and where
\begin{equation}
 (\tilde{Q}^{-1}) _{\mu \nu}{}^{\rho \sigma} \tilde{Q} _{\rho \sigma}{}^{\alpha \beta}=\delta_{\mu \nu}{}^{\alpha \beta}\, , 
\end{equation}
so that $\tilde{Q}^{-1}$ is the inverse tensor of $\tilde{Q}$, as described in the previous equation, and $\delta_{\mu \nu}{}^{\rho \sigma}=\delta_{[\mu}{}^{[\rho} \delta_{\nu]}{}^{\sigma]}$. We are also defining $\alpha_{1,s,1} \equiv \alpha_{1,s}$ and $\alpha_{2,s,1} \equiv \alpha_{2,s}$. Finding such inverse tensor is generically a rather challenging task, but it is a more manageable one when we restrict ourselves to backgrounds with enough symmetry. In the case at hand, since we are considering a pure AdS space with $R^{\mu \nu}{}_{\rho \sigma}=-2/\tilde{L}^2 \delta^{\mu \nu}{}_{\rho \sigma}$, $\tilde{L}=L/\sqrt{f_\infty}$, we have
\begin{equation}
Q^{\mu \nu}{}_{\rho \sigma}=\tilde{Q}^{\mu \nu}{}_{\rho \sigma}= \alpha_{\rm eff} \delta^{\mu \nu}{}_{\rho \sigma} \, , \quad ( \tilde{Q}^{-1})_{\mu \nu}{}^{\rho \sigma}=\frac{1}{\alpha_{\rm eff}} \delta^{\mu \nu}{}_{\rho \sigma}\,.
\end{equation}
Consequently, the coefficient of $F^2$ in \eqref{eq:dualth} turns out to be $1/\alpha_{\rm eff}$. This implies that the net effect of the higher-order terms is the renormalization of the gauge coupling constant, producing in turn the central charge
\begin{equation}
C^{\rm EQG}_J =\frac{C_J^{\rm EM}}{\alpha_{\rm eff}}\, , \quad C_J^{\rm EM}=\frac{\Gamma(d)}{\Gamma(d/2-1)} \frac{\ell_\ast^2 \tilde{L}^{d-3}}{4 \pi^{d/2+1} G}\,,
\label{eq:cjgen}
\end{equation}
being $C_J^{\rm EM}$ the Einstein-Maxwell central charge. On the other hand, the computation of $a_2$ requires the knowledge of the inverse tensor $\tilde{Q}^{-1}$ on a shock-wave background ---see \cite{Hofman:2008ar,Cano:2022ord} for more details in this computation--- given by the metric 
\begin{align}
\mathrm{d}s^2&=\frac{\tilde{L}^2}{u^2}\bigg[ \delta(y^+) \mathcal{W}(y^i,u) \left ( \mathrm{d} y^+ \right)^2-\mathrm{d} y^+\mathrm{d} y^-+ \sum_{j=1}^{d-2} \left ( \mathrm{d} y^j \right)^2 +\mathrm{d}u^2\bigg]\,, \\ \mathcal{W}(y^i,u)&= \frac{\mathcal{W}_0 u^d}{\left (u^2+ \sum_{j=1}^{d-2}(y^j-y_0^j)^2 \right)^{d-1} }\, , \quad y_0^j \in \mathbb{R}\,.
\end{align}
This shock-wave background satisfies that $R_{\mu \nu}=-d/\tilde L^2 g_{\mu \nu}$ and, being a Brinkmann spacetime, the square of its Weyl tensor vanishes, \emph{i.e.} $W_{\mu \nu \rho \sigma} W^{\rho \sigma \alpha \beta}=0$. Taking into account this properties, it can be seen that
\begin{equation}
\tilde{Q}_{\mu \nu}{}^{\rho \sigma}= \alpha_{\rm eff} \delta_{\mu \nu}{}^{\rho \sigma}  -\frac{\beta_{\rm eff}}{f_\infty (d-1)(d-2)} W_{\mu \nu}{}^{\rho \sigma}\, , \quad  ( \tilde{Q}^{-1})_{\mu \nu}{}^{\rho \sigma}= \frac{1}{\alpha_{\rm eff}} \delta_{\mu \nu}{}^{\rho \sigma} + \frac{\beta_{\rm eff}}{f_\infty (d-1)(d-2) \alpha_{\rm eff}^2} W_{\mu \nu}{}^{\rho \sigma} \, ,
\end{equation}
where $\beta_{\rm eff}$ is the parameter introduced in \req{betaeff}.
We identify this result as formally equivalent to that of Eq.\,(4.46) of \cite{Cano:2022ord}, obtained in the context of the four-derivative theory \eqref{EQf}, upon exchange of $-(2(2d-1)(d-1)\alpha_1+2(d-1) \alpha_2)\mapsto \beta_{\rm eff}/f_\infty$. Hence the coefficient $a_2$ associated to \eqref{eq:geneqg} will be that of Eq.\,\eqref{eq:a2}, after making the aforementioned substitution, namely,
\begin{equation}
a_2^{\rm EQG} =\frac{d \beta_{\rm eff} }{(d-2) \alpha_{\rm eff}}\,.
\label{eq:a2gen}
\end{equation}
Therefore, taking into account Eqs.\,\eqref{eq:cjgen} and \eqref{eq:a2gen}, we notice that the entanglement entropy can be rewritten as
\begin{equation}
S_{\rm \ssc EE}=\frac{\tilde{L}^{d-1}V_{\mathbb{H}^{d-1}} }{4 G}\left[1-\frac{2(d-1)}{d-3}\lambda f_{\infty}\right] +\frac{\Gamma(d/2-1) \pi^{d/2+1}  V_{\mathbb{H}^{d-1}} }{\Gamma(d)} C^{\rm EQG}_J   \left (\frac{(d-2)^2}{d-1}+ \frac{(d-2)^3 a^{\rm EQG}_2}{d (d-1)^2 } \right)(\mu R)^2+\mathcal{O}(\mu^4)\,.
\end{equation}
The regularized volume of the unit hyperbolic space is given by \cite{CHM} $V_{\mathbb{H}^{d-1}}=\nu_{d-1}/(4 \pi) \Omega^{d-1}$, where $\Omega^{d-1}$ is the volume of the unit sphere $\mathbb{S}^{d-1}$ and $\nu_{d-1}$ is defined as in \eqref{eq:nuentro}. We then arrive at our final result
\begin{equation}
\frac{S_{\rm \ssc EE}(\mu)}{\nu_{d-1}}=a^\star_{\rm GB} +\frac{ \pi^{d}  }{(d-1)^2\Gamma(d-2)} C^{\rm EQG}_J   \left [1+ \frac{(d-2) a^{\rm EQG}_2}{d (d-1) } \right](\mu R)^2+\mathcal{O}(\mu^4)\,,
\label{eq:resfinapp}
\end{equation}
where we have introduce the $a^\star$ charge of Gauss-Bonnet theory, presented in Eq.\,\eqref{eq:astargb}. Eq.\,\eqref{eq:resfinapp} is then a realization of Eq.\,\eqref{Suniv} for our infinite family of theories and we  fulfil  the goal of the appendix. 

As a final comment, one may wonder about the effect of including arbitrary pure-gravity quasitopological higher-order terms \cite{Quasi,Quasi2,Dehghani:2011vu,Ahmed:2017jod,Cisterna:2017umf,Bueno:2019ycr}  into the action \eqref{eq:geneqg}. Given the structure and derivation of Eq.\,\eqref{eq:resfinapp}, we expect such pure-gravity terms to simply produce a renormalization of the constant $f_\infty$, while leaving Eq.\,\eqref{eq:resfinapp} invariant.

\bibliographystyle{apsrev4-1} 
\vspace{1cm}
\bibliography{Gravities} 

\end{document}